\documentclass[12pt,preprint]{aastex}
\usepackage{graphicx}

\def\arcsec{\hbox{$^{\prime\prime}$}}
\def\arcdeg{\mbox{$^\circ$}}%

\newcommand{\apg}{\gtrsim}
\newcommand{\apl}{\lesssim}

\newcommand{\etal}{et al.}
\newcommand{\hI}{\mbox{${\rm H\,I}$}}
\newcommand{\nI}{\mbox{${\rm N\,I}$}}
\newcommand{\kms}{\mbox{km\ s${^{-1}}$}}
\newcommand{\lya}{\mbox{${\rm Ly}\alpha$}}
\newcommand{\lyb}{\mbox{${\rm Ly}\beta$}}

\begin{document}

\lefthead{Chen \etal}
\righthead{}

\slugcomment{Accepted for Publication in the Astrophysical Journal}

\title{SUPER STAR CLUSTER NGC\,1705-1: A LOCAL ANALOGUE TO THE BIRTHSITE OF 
LONG-DURATION $\gamma$-RAY BURSTS}
\author{
HSIAO-WEN CHEN\altaffilmark{1}, 
JASON X.\ PROCHASKA\altaffilmark{2}, and 
JOSHUA S.\ BLOOM\altaffilmark{3,4}
}


\begin{abstract}

Recent observations suggest that global properties of the host
galaxies for long-duration $\gamma$-ray bursts (GRBs) are particularly
well-suited for creating the massive star progenitors of these GRBs.
Motivated by the hypothesis that massive young star clusters located
in metal-poor, low-metallicity galaxies are a natural birthsite for
GRB progenitors, we present a comparison study of the ISM properties
along the sightline toward the super star cluster, NGC\,1705-1, and
those in distant GRB hosts.  Using the same set of metal transitions
in the UV and applying known ISM structures in NGC\,1705, we find that
NGC\,1705-1 resembles distant GRB host galaxies in its high neutral
gas column, low molecular gas fraction, low metallicity,
$\alpha$-element enhancement,and low dust depletion.  The lack of
molecular gas is due to the enhanced UV radiation field in the
starburst environment, consistent with the expectations for GRB
progenitors.  In addition, the known presence of dense neutral gas
clouds at $r\apl 500$ pc from NGC\,1705-1 provides a natural reservoir
of C$^+$, Si$^+$, and Fe$^+$ ions that may be subsequently excited by
the afterglow UV radiation field to produce excited lines commonly
seen in GRB host ISM.  We further argue that the apparent offset in
the velocity profiles of low- and high-ionization transitions from
absorption-line observations alone already offers important clues for
related starburst episodes in GRB host galaxies. Our study shows that
a statistical comparison between the ISM around star clusters and
high-redshift GRB progenitors is important for identifying the key
physical parameters that facilitate the formation of GRBs.

\end{abstract}

\altaffiltext{1}{Department of Astronomy \& Astrophysics, University of
Chicago, Chicago, IL 60637, {\tt hchen@oddjob.uchicago.edu}}

\altaffiltext{2}{UCO/Lick Observatory; University of California, Santa
  Cruz, Santa Cruz, CA 95064, {\tt xavier@ucolick.org}}

\altaffiltext{3}{Department of Astronomy, 601 Campbell Hall, University of 
California, Berkeley, CA 94720 {\tt jbloom@astron.berkeley.edu}}

\altaffiltext{4}{Sloan Research Fellow}


\section{INTRODUCTION}

\setcounter{footnote}{0}

  Mounting evidence has lent strong support for the origin of
long-duration gamma-ray bursts (GRBs) in the death of massive stars
(see Woosley \& Bloom 2006 for a review; but also see Fynbo \etal\
2006a, Gal-Yam \etal\ 2006, \& Della Valle \etal\ 2006 for recent
discoveries).  Specifically, a direct association between GRBs and
Type Ib/c supernovae (SNe) is established for GRBs 980425, 030329, and
031203 (Patat \etal\ 2001; Hjorth \etal\ 2003; Stanek \etal\ 2003;
Malesani \etal\ 2004).  These SNe of Type Ic show no hydrogen or
strong SiII transitions in their spectra, and have never been seen in
elliptical galaxies (Filippenko 1997).  While it is conceivable that
all long-duration GRBs occur contemporaneously with a Type Ic SN,
their low event rate indicates that only a small fraction ($< 0.1$\%)
of Type Ic SNe are accompanied by a GRB phase (van Putten 2004;
Podsiadlowski \etal\ 2004; Soderberg \etal\ 2006).  For stars to lose
their hydrogen envelope prior to the core collapse and form a GRB,
current theoretical models have homed in on progenitors with initial
stellar mass of $M=25-30\ M_\odot$, fast rotation speed, and low
metallicity $Z\apl 10\%\ Z_\odot$ (Woosley \& Heger 2007).

  These GRBs are therefore expected to originate preferentially in
metal-poor environment and/or underluminous galaxies.  Observations of
five nearby ($z_{\rm GRB}<0.17$) host galaxies indeed show that the
host galaxies tend to be underluminous and that two of these galaxies
also have an integrated H\,II-region abundance below what is expected
from a nominal luminosity-metallicity relation of field galaxies
(e.g.\ Prochaska \etal\ 2004; Modjaz \etal\ 2006).  At the same time,
chemical abundance studies of high-redshift ($z>1$) hosts have been
carried out based on absorption lines identified in early-time
afterglow spectra (e.g.\ Fynbo \etal\ 2006b; Savaglio 2006; Prochaska
\etal\ 2007), where the emission lines of an H\,II region become too
faint to be detected.  Abundance measurements of the cold interstellar
medium (ISM) based on absorption-line studies, despite representing
only a small volume along a pencil beam from the afterglow, are less
subject to uncertainties in the ionization fraction or temperature of
the gas, and therefore offer a more accurate assessment of the
metallicity.  Comprehensive studies of afterglow absorption-line
spectra have also allowed us to resolve gas at $\sim 100$ pc from the
progenitor stars, providing direct constraints on the ISM properties
of the progenitor environment (e.g.\ Prochaska, Chen, \& Bloom 2006;
Chen \etal\ 2007; Vreeswijk \etal\ 2007).

  However, our understanding of the progenitor environment is far from
clear.  For example, roughly 50\% of GRB afterglows are located behind
a large column of neutral gas with $\log\,N(\hI)=21.5-22.5$ (Jakobsson
\etal\ 2006) but no trace of molecular gas to very low limits $f_{\rm
H_2} = 2\,N({\rm H_2})/[N(\hI)+2\,N({\rm H_2})]\apl 10^{-6}$
(Tumlinson \etal\ 2007).  Using a sample of 16 afterglow spectra,
Prochaska \etal\ (2007) showed that GRB hosts have sub-solar chemical
abundances and an $\alpha$-element enhancement relative to Fe in the
ISM, but also that a non-negligible fraction exhibit $>10$\% solar
abundances.  In addition, these authors also showed that even with a
large neutral gas column, there can be little dust extinction along
the sightlines in the host ISM toward the afterglows (see also
Starling \etal\ 2007).  The lack of H$_2$ can, in principle, be
explained by a high destruction rate due to the enhanced
photo-ionization radiation field in the star-forming cloud that hosts
the GRB progenitor and by a low formation rate inferred from the
observed low metallicity.  But at high enough gas density ($n_{\rm
H}>10^3$ cm$^{-3}$), self-shielding is expected to be effective and
will promote H$_2$ survival (e.g.\ Tumlinson \etal\ 2007).

  Super star clusters provide an interesting local analogue for
understanding the formation of GRBs for two main reasons.  First,
Wolf-Rayet stars, a popular candidate for GRB progenitors (e.g.\
Woosely \& Bloom 2006), are often found in young star clusters (see
Crowther 2006 for references).  Second, star clusters contain a large
number of massive stars in a small volume and are expected to have on
average a higher rate of stellar collision and mass transfer between
individual stars for nurturing the formation of a GRB.  Various
authors have proposed that dense stellar environments facilitate
stellar coalescence for forming blackholes that power the GRBs (e.g.\
Hansen \& Murali 1998; Efremov 2001).  In addition, GRBs have been
considered as a possible source for powering giant HI supperbubble
shells found in nearby galaxies (Efremov, Elmegreen, \& Hodge 1998;
Loeb \& Perna 1998) A direct comparison between absorption-line
properties in the ISM surrounding super star clusters and those
identified in GRB host galaxies may therefore provide important
calibrations for interpreting afterglow absorption-line observations.

  To gain further insights for interpreting the afterglow
absorption-line spectra of high-redshift GRBs, we have searched for
ultraviolet echelle spectra of nearby super star clusters in the
Hubble Space Telescope (HST) Data Archive.  NGC\,1705-1, a super star
cluster sitting at the center of a blue compact dwarf galaxy NGC\,1705
at redshift $z=0.002$, is the only star cluster with high quality
echelle spectra from the Space Telescope Imaging Spectrograph (STIS).
The galaxy is a well known starburst galaxy that exhibits evidence of
galactic scale gas outflows.  It has a total luminosity of $\sim 0.01\
L_*$ and an H\,II region abundance of $\approx$ 35\% solar (e.g.\ Lee
\& Skillman 2004).  A comparison between the on-going star formation
rate and $B$-band luminosity indicates a formation time scale that is
much shorter than normal irregular galaxies and comparable to what is
known for GRB host galaxies at high redshift (e.g.\ Christensen,
Hjorth, \& Gorosabel 2004).  The spectra of the super star cluster
were originally taken for studying the physical conditions of the
galactic winds driven by starburst outflows (see e.g.\ V\'azquez
\etal\ 2004).  Here we re-visit this data set and present new column
density measurements of various low-ionization species.  We compare
the ISM properties of the dense star cluster with those of
high-redshift GRB host galaxies.

  This paper is organized as the following.  In Section 2, we review
known ISM properties integrated over the entire galaxy NGC\,1705 and
ISM properties that are local to the super star cluster NGC\,1705-1.
In Section 3, we present our absorption-line measurements of the cold
gas along the line of sight toward NGC\,1705-1.  In Section 4, we
compare the ISM properties of NGC\,1705-1 with measurements for
high-redshift GRB hosts.  We describe the value of collecting echelle
spectra of nearby supper star clusters using future space UV
facilities such as the Cosmic Origin Spectrograph, for understanding
the GRB progenitor environment at high redshift.

\section{REVIEW OF NGC\,1705}

  NGC\,1705 is a well-known starburst galaxy at $z=0.00209$ that has
been studied extensively over a broad spectral window, from
ultraviolet through radio frequencies (e.g.\ Lamb \etal\ 1985;
Calzetti \etal\ 1994; Meurer \etal\ 1998; Lee \& Skillman 2004).  This
blue irregular galaxy hosts a super star cluster, NGC\,1705-1 (Figure
1), that is among the brightest star clusters in the local universe
and contributes roughly 40\% of the total UV light of the galaxy
(Meurer \etal\ 1992).  Spatially resolved strong emission lines
indicate that the warm ISM in the H\,II region can be charaterized
with an electron temperature of $T_e=(1.1-2)\times 10^4$ K (Lee \&
Skillman 2004).  One of the most notable features in the ISM of
NGC\,1705 is the multiple super bubble shells, indicating gas outflow
driven by the last episode of starburst roughly $10^7$ years ago that
also formed most of the stars ($\sim\,10^5\ M_\odot$) seen today in
NGC\,1705-1 (Meurer \etal\ 1992).  The galaxy also exhibits very
little intrinsic reddening (Calzetti \etal\ 1994) and shows no sign of
molecular gas to very sensitive levels (Greve \etal\ 1996; Hoopes
\etal\ 2004).

  Moderate-to-high resolution UV spectra of NGC\,1705-1 from different
space facilities are published in the literature.  While most of the
absorption lines identified in the spectra are of ISM origin (York
\etal\ 1990), detailed stellar population analysis based on a few
stellar photospheric lines have placed strong constraints on the age,
metallicity, and IMF of the cluster (V\'azquez \etal\ 2004).  The ISM
absorption lines exhibit two strong components with respective
blue-shifted velocity of $v\approx -43$ and $v\approx -77$ \kms\
(Heckman \etal\ 2001; V\'azquez \etal\ 2004).  The low velocity
component is believed to trace the cold, neutral ISM, while the
high-velocity component is interpreted to originate in a gas outflow.
Abundance measurements based on low ions such as N$^0$, S$^+$ and
Fe$^+$ support the scenario that the cold neutral ISM has been
enriched primarily by Type II SNe that occurred in the star cluster
roughly $10^7$ yr ago (Sahu \& Blades 1997; Sahu 1998).  In addition,
the observed kinematic profile and the absorption strength of the
O\,VI doublet are consistent with the expectations of superwinds
undergoing radiative cooling (Heckman \etal\ 2001).

  A summary of known properties of galaxy NGC\,1705 and the super star
cluster NGC\,1705-1 is presented in Table 1.  We note that ISM
absorption-line measurements in the table are based primarily on
previous FUSE observations (e.g.\ Heckman \etal\ 2001; Hoopes \etal\ 2004).

\section{ISM PROPERTIES OF NGC\,1705-1}

\subsection{STIS Echelle Spectra}

  We have retrieved all available STIS echelle spectra of NGC\,1705-1
in the HST Data Archive.  The observations were carried out using a
$0.2\arcsec\times 0.2\arcsec$ slit and three different grating set-ups
for a total exposure time of 28,569 s (PID=8297).  Individual echelle
spectra were reduced, extracted, and calibrated using standard
pipeline techniques.  To form a final spectrum for absorption line
studies, we calculated a weighted average of individual spectra and
the corresponding 1-$\sigma$ error array per grating set-up with the
weighting factor determined by the inverse variance.  We normalized
the mean spectrum with a best-fit, low-order polynomial continuum.
The combined and normalized spectra have a spectral resolution of
${\rm FWHM}\approx 7$ \kms, covering wavelength ranges of
$\lambda=1150-1708$ \AA\ with the E140M grating at signal-to-noise
ratio ${\rm S/N}\approx 12$ and $\lambda=1700-3060$ \AA\ with the
E230M grating at ${\rm S/N}\approx 7$ per resolution element.

\subsection{Column Density Measurements}

  Absorption lines of heavy ions identified in the ISM of NGC\,1705
exhibit complex absorption profiles.  In Figure 2, we present in the
three left panels low ionization species, and in the right column
high-ionization species.  The zero relative velocity ($v=0$)
corresponds to the systematic redshift of NGC\,1705 $z=0.00209$,
measured in H\,I 21cm (Meurer \etal\ 1992) and in stellar photospheric
transitions (V\'azquez \etal\ 2004).  The spectra include absorption
features from the Milky Way and a foreground high-velocity cloud (Sahu
\& Blades 1997) which occasionally blend with features from NGC\,1705.
These `contaminating' features have been identified and are indicated
with dotted lines in the panels.  In addition, stellar photospheric
features are apparent in transitions such as C\,IV and Si\,IV, as a
broad, shallow component extending from $v=-500$\kms\ through $v=+300$
\kms.

  The low- and high-ionization species in the ISM of NGC\,1705 exhibit
distinct kinematic signatures.  The low-ion resonance lines, (which
excludes excited C$^+$ transition, C\,II* 1335), show two narrow
components at $v=-20$ and $-40$ \kms, while the high-ionization
elements are broad and show velocity centroids systematically
blue-shifted to $v=-70$ \kms.  These high ions share the same
kinematic feature as the O\,VI absorption doublet discussed in Heckman
\etal\ (2001).

  We measure the column density of each ion using the apparent optical
depth method (Savage \& Sembach 1991) over a velocity interval of
$v=-100$ to $+40$ \kms\ for low-ionization transitions and
$v=-250$ to $+100$ \kms\ for high-ionization transitions.  One
exception is for the S\,II 1259 transition, where the line is
contaminated by the Si\,II 1260 transition from the foreground
high-velocity cloud.  We determine $N({\rm S\,II})$ using the VPFIT
software package, including multiple components.  The measurements are
summarized in Table 2.

  The H\,I \lya\ transition of NGC\,1705 is present in the STIS
spectrum.  It is saturated and blended with absorption from the Milky
Way (Figure 3).  We note, however, that the $N(\hI)$ value of
NGC\,1705 is well constrained by the red damping wing.  We have
performed a Voigt profile analysis and found $\log\,N(\hI)=20.05\pm
0.15$ for NGC\,1705 and $\log\,N(\hI)=19.7$ for the Milky Way. This is
consistent with the finding of Sahu (1998) and Heckman \etal\ (2001)
based on HST GHRS and FUSE observations, respectively.  The best-fit
model profile combining absorption from the Milky Way and NGC\,1705 is
presented as the red curve in Figure 3.  The total H\,I gas column
density observed along the sightline toward NGC\,1705-1 is only about
10\% of what is observed in H\,I 21 cm emission (Meurer \etal\ 1992),
indicating that the super cluster is located near the front edge of
the galaxy and the sightline only probes a fraction of the (presumably
outer) gaseous disk.  

\subsection{Chemical Composition of the Cold ISM}

  Based on the column density measurements presented in Table 2, we
derive an iron abundance of $[{\rm Fe}/{\rm H}]=-1.1 \pm 0.2$ and
$[{\rm S}/{\rm H}]=-0.6 \pm 0.1$ with respect to the solar values from
Asplund, Grevesse, \& Sauval (2005).  The sulfer abundance in the cold
neutral medium is consistent with the oxygen abundance determined for
the H\,II region, $[{\rm O}/{\rm H}]=-0.5\pm 0.05$ (Table
1)\footnote{We note the low oxygen abundance $[{\rm O}/{\rm H}]$
determined by Heckman \etal\ (2001) based on O\,I absorption lines
observed in FUSE spectra.  Given the observed column density, however,
we expect the line to be saturated.  Therefore, the reported value
should be treated as a lower limit.}.

  The large ratio of $[{\rm S}/{\rm Fe}]=+0.5 \pm 0.2$ indicates a an
$\alpha$-element enhancement, which is qualitatively consistent with a
chemical enrichment history dominated by Type II supernovae and
little/no reddening observed for the ISM of NGC\,1705 (Calzetti \etal\
1994).  Although dust depletion can be substantial even when reddening
is negligible (e.g.\ Savaglio 2006; Prochaska \etal\ 2007), the
significant underabundance of nitrogen, $[{\rm N}/{\rm H}]=-2.1 \pm
0.1$, from Heckman \etal\ (2001) indicates that more evolved,
intermediate-mass stars have not formed to produce abundant nitrogen
and is consistent with a chemical enrichment history dominated by Type
II supernovae (e.g.\ Henry et al.\ 2000).

  In contrast with the GRB host environment observed at high redshift,
we do not observe excited ions (e.g.\ strong C\,II* 1335 transition)
for the neutral gas in the ISM toward NGC\,1705-1.  Instead, the
observed C\,II* 1335 transition is blueshifted by $\approx 30$ \kms\
from the neutral species (see discussion in the following section).
The lack of excited C$^+$ in the neutral gas implies a warm neutral
medium and/or low dust content (Wolfe, Prochaska, \& Gawiser 2003).

\subsection{Gas Density and Temperature of the Ionized Gas}

  As noted earlier, high-ionization transitions such as C\,IV 1548,
1550, Si\,III 1206, Si\,IV 1393, 1402, and Al\,II 1854, 1862 are
observed to be blue-shifted at $v\approx -77$ \kms\ in the STIS
echelle spectra (Figure 2).  These coincide in velocity space with the
O\,VI absorption doublet reported by Heckman \etal\ (2001).  In
particular, we note the presence of strong C\,II* at the same velocity
offset, where little Fe\,II absorption has been detected.  This is
similar to what is observed in the ISM toward HD\,192639
(Sonnentrucker \etal\ 2002), whose origin is interpreted to be in
shocks produced by expanding supperbubble shells or stellar winds.

  The broad line widths seen in unsaturated lines such as C\,II* 1335
and Al\,III 1854 are suggestive of a collisional ionization origin.
Adopting the observed S\,III and S\,IV transitions that share the same
kinematic features with the O\,VI absorber (Heckman \etal\ 2001), we
find $\log\,N({\rm S\,III})-\log\,N({\rm S\,IV})<0.57$ and constrain
the gas temperature at $T> 6\times 10^4$ K under a collisional
ionization equilibrium.  Further constraints for the gas temperature
can be obtained by comparing the line width of C\,II* 1335 and Al\,III
1854.  We measure a Doppler parameter of $b_{\rm C\,II*}=22.5\pm 1.2$
\kms\ for C$^+$ and $b_{\rm Al\,III}=17.73\pm 5.4$ \kms\ for
Al$^{2+}$ from the STIS spectra.  Assuming that these ions originate at
the same location, we further infer a gas temperature of $T\approx
2.5\times 10^5$ K and a turbulent velocity of $b_0\approx 12.6$ \kms.

  Under a collisional excitation scenario, we can also constrain the
density of ionized gas based on the relative strengths between
resonance absorption and absorption from excited states.  Given that
the C\,II 1334 transition is fully saturated, we obtain the contraint
using the relative abundance between excited Si$^+$ and ground-state
Si$^+$.  The high-velocity component at $v=-77$ \kms\ is moderately
resolved from the warm and cold gas components.  We perform a
multiple-component Voigt profie analysis and measure $\log\,N({\rm
Si\,II})=13.8\pm 0.1$ and $b=8.7\pm 0.9$ \kms\ for the high-velocity
component.  The lack of Si\,II* yields $N({\rm Si\,II*})/N({\rm
Si\,II}) < 0.04$, leading to an electron density $n_e\apl 50$
cm$^{-3}$ for $T_e \sim 10^5$ K.

\section{DISCUSSION}

  Recent observations of GRB host environments have yielded two main
results.  First, the ISM of long-duration GRBs appear to be
essentially devoid of molecular gas, despite exhibiting strong
neutral-hydrogen absorption columns.  Second, these GRBs occur
preferentially in metal-poor, low-metallicity, low-luminosity,
actively star forming galaxies.  Together these suggest that global
properties of the hosts are particularly well-suited for the creation
of the massive star progenitors thought to give rise to GRBs.  If
true, the study of local analogues of the sorts of environments
amenable to GRB progenitor birth should help in understanding the
nature of distant GRB birthsites.

  NGC\,1705 is a well known starburst galaxy in the nearby universe,
with special interest because of the clear presence of substantial
O$^{5+}$ in a starburst-driven outflow.  We have re-visited and
presented new measurements of the ISM properties along the line of
sight toward the super star cluster, NGC\,1705-1, using archived by
unanalyzed STIS echelle spectra.  The study is motivated by the
hypothesis that massive young star clusters offer a birthsite for the
progenitor stars of high-redshift GRBs, because of the dense stellar
environment (e.g.\ Hansen \& Murali 1998; Efremov 2001).  The
available STIS UV echelle spectra of the cluster show numerous ionic
transitions with a complex velocity structure, and allow us to use the
same set of metal absorption lines for a direct comparison between
known properties of the ISM near the progenitors of high-redshift GRBs
and the ISM properties around NGC\,1705-1.  Here we discuss the
insights we gain from the UV echelle spectra of nearby super star
clusters for interpreting the afterglow absorption-line studies of
distant GRB hosts.

\subsection{Gaseous Structures of the ISM near NGC\,1705-1}

  The complex velocity structure displayed in the absorption-line
profiles of different ionization species indicate a multi-phase ISM
near the super star cluster NGC\,1705-1.  Based on FUSE observations
of the starburst, Heckman \etal\ (2001) presented a comprehensive
study of the thermal state of the ISM.  These authors identified (1) a
neutral medium as demonstrated by H\,I, O\,I, Si\,II, and Fe\,II
absorption, (2) a warm photoionized phase as demonstrated by the
H$\alpha$ emission nebula (Meurer \etal\ 1992; Marlowe \etal\ 1995)
and N\,II and S\,III absorption, and (3) coronal gas as demonstrated
by the O\,VI absorption doublet.  These results are well supported by
our analysis of the STIS echelle observations.

  Given the small velocity offset of the neutral gas ($v\approx 43$
\kms) with respect to the nebular lines, Heckman \etal\ propose that
the neutral gaseous clouds represent neutral ISM that has been
engulfed by the expanding superbubble.  Under this scenario, the
distance of the neutral gas must be $d_{\rm neutral} \apl 500$ pc, the
observed size of the superbubble, and the gas density is required to
be $n_{\rm neutral} \apg 8\ {\rm cm}^{-3}$ for effective self-shieding
from the ionizing radiation of the star cluster.  Together these
constrain the cloud to be $\sim 10$ pc for producing the observed H\,I
column density.

  Given the large observed column density of O$^{+5}$ at $v\approx 77$ \kms,
Heckman \etal\ also concluded that the coronal gas cannot originate
from behind the shocks induced by the expanding superbubble, but
rather blowout gas (super winds) that are undergoing radiative cooling
between photo-ionized fragments of the shells.  In this picture, the
ISM around NGC\,1705-1 has been heated by both the expansion of super
bubbles and the UV radiation of the central star cluster.  The STIS
echelle observations allow us to constrain the gas temperature at
$T_{\rm coronal} \sim 10^5$ K and a bulk motion of of $b_0\approx
12.6$ \kms.  

\subsection{Comparisons with the ISM Properties of High-redshift GRB Hosts}
 
  The primary objectives of the present study are (1) to explore an
alternative scenario in which young super star clusters provide a
likely birthsite for long-duration GRBs, and (2) to examine whether
available UV echelle spectra of a nearby super star clusters offer
additional insights for interpreting afterglow absorption-line studies
of high-redshift bursts. Our current understanding of the host ISM
around high-redshift bursts can be summarized as the following.

  First, Prochaska, Chen, \& Bloom (2006) argued that the large
neutral gas column observed along an afterglow sightline is located at
$r=100-1000$ pc from the GRB progenitor star, in order for Mg$^0$ to
survive the intense UV radiation field.  The associated heavy ions
such as C$^+$, Si$^+$, and Fe$^+$ in the neutral gas clouds would be
subsequently excited by the afterglow UV photons for producing the
observed absorption features from their excited states (confirmed by
Vreeswijk \etal 2007 for GRB\,060418).  The neutral gas clouds in the
ISM model of NGC\,1705-1 from Heckman \etal\ (2001) offers a natural
candidate for the presence of dense neutral clouds in the vicinity of
a GRB progenitor at high redshift\footnote{We note that the STIS
echelle spectra presented in this paper were obtained using a small
aperture $0.2\arcsec\times 0.2\arcsec$ that was centered on
NGC\,1705-1, in contrast to the $30\arcsec$ aperture available for the
FUSE observations.  The agreement between our respective $N(\hI)$ and
$N({\rm Fe\,II})$ measurements, therefore, confirms that the neutral
gas close to the super star cluster dominates the absorption column
observed in FUSE.}.  Note that the observed $N(\hI)$ along the
sightline toward NGC\,1705-1 is more similar to strong Lyman limit
systems of $N(\hI)=19-20$ observed in the hosts of GRBs\,021004,
050908, and 060526 (Jakobsson \etal\ 2006).  Higher column density
absorbers would imply either a smaller cloud size or higher neutral
gas density.  Nonetheless, should a GRB occur in NGC\,1705-1, the
neutral gas at $r\sim 500$ pc would still remain neutral under the
afterglow radiation field, but would be excited to produce the
commonly observed transitions from excited ions.

  In addition, the ISM of NGC\,1705 exhibits little/no trace of
molecular gas, neither CO nor H$_2$, despite the recent starburst
episode $\sim\,10$ Myr ago and on-going star formation $\sim
0.06\,M_\odot$ yr$^{-1}$.  NGC\,1705 resembles the Small Magellanic
Cloud both in size and in the total metal content, but has on average
a substantially smaller molecular fraction: $f_{\rm H_2}<5.2\times
10^{-7}$ for NGC\,1705-1 (Table 1) versus $\langle f_{\rm H_2} \rangle
\lesssim 0.01$ near star-forming regions in SMC (Tumlinson \etal\
2002).  The super star cluster NGC\,1705-1 near the center of
NGC\,1705 is among the brightest star clusters in the local universe
(Meurer \etal\ 1995), containing massive stars of $> 30\ M_\odot$ and
$\sim 10$ Myr old.  

  The lack of molecules in the ISM around NGC\,1705-1 is reminiscent
of what is observed in the high column density clouds surrounding GRB
progenitors (Hatsukade \etal\ 2007; Tumlinson \etal\ 2007).  The low
molecular fraction and dust extinction for moderately enriched ISM and
high gas density appear to be remarkably similar, implying that the
lack of molecules and dust may be due to the enhanced UV radiation
from massive stars in the star-forming region that hosts the GRB
progenitor (e.g.\ Hoopes \etal\ 2004; Tumlinson \etal\ 2007).

  We further note that similar to what is observed in the afterglow
spectra of high-redshift GRBs, low-ionization species in NGC\,1705-1
are more confined in a narrow veolicty interval whereas
high-ionization transitions exhibit larger velocity spread.  The
apparent offset in the velocity profiles of low-ionization (such as
Fe\,II) and high-ionization transitions (such as C\,IV and Si\,IV)
observed toward NGC\,1705-1 (Figure 2) is now understood as due to
starburst driven outflows, based on extensive imaging and spectroscopy
studies.  It suggests that important clues for possible starburst
nature in GRB host galaxies may already be learned based on afterlow
spectra alone.  The is particularly valuable because the host galaxies
of distant GRBs are faint and challenging to be found in emission.  Of
the six GRB host studied in Chen \etal\ (2007), we identify velocity
offsets between Si\,II and C\,IV/Si\,IV over the range from $v=-65$ to
$v=-200$ \kms, along the sightlines toward GRBs\,021004, 050730,
050908, and 060418.

  Finally, we present in Table 3 a direct comparison of different ISM
properties between NGC\,1705-1 and GRB hosts.  The table shows that
the two sources occupy a comparable parameter space.  In particular,
absorption-line studies of GRB host galaxies show that the host ISM of
high-redshift GRBs exhibits clear $\alpha$-element enhancement with
little dust extinction/depletion along the sightlines toward the
afterglows.

  The similarity in the ISM properties shows that this scenario for
long-duration GRBs to originate in massive star clusters is at least
viable.  The relatively low nitrogen abundance also signifies the
young age of the progenitor star clusters, because nitrogen is
produced primarily in intermediate-mass AGB stars (e.g.\ Prochaska
\etal\ 2007).  NGC\,1705-1 is unique among known super star clusters
locally, because of its young age and massive stellar content.
Conclusive evidence requires a large UV spectroscopic sample of nearby
super star clusters.  These data are necessary for a complete
statistical analysis of their ISM properties to be compared with
afterglow absorption-line studies.  For example, a comparison of the
$N(\hI)$ distribution and the size distribution of neutral clouds in
the vicinity of young star clusters may offer important support, or
otherwise, for the star cluster origin of long-duration GRBs.  We
expect that a direct comparison with known statistical properties of
high-redshift GRB hosts will allow us to identify the key physical
parameters that facilitate the formation of GRBs. It is worthwhile to
note also the direct evidence of inflow gas around NGC\,1705, as
evidenced by the redshifted components seen in C\,IV and Si\,IV.  A
larger sample may be useful for studying gas acretion around galaxies.
Future spectroscopic observations of nearby super star clusters using
the Cosmic Origin Spectrograph to be installed on the HST are
necessary.

\acknowledgments
 
  We thank J. Gallagher for a coffee discussion that sparked this
analysis and D.\ Welty for his valuable input on HD\,192639.  This
research has made use of the NASA/IPAC Extragalactic Database (NED)
which is operated by the Jet Propulsion Laboratory, California
Institute of Technology, under contract with the National Aeronautics
and Space Administration.  H.-W.C., JXP, and JSB acknowledge support
from NASA grant NNG05GF55G.  H.-W.C. acknowledges partial support from
an NSF grant AST-0607510.


\begin{figure}
\begin{center}
\includegraphics[scale=0.7, angle=0]{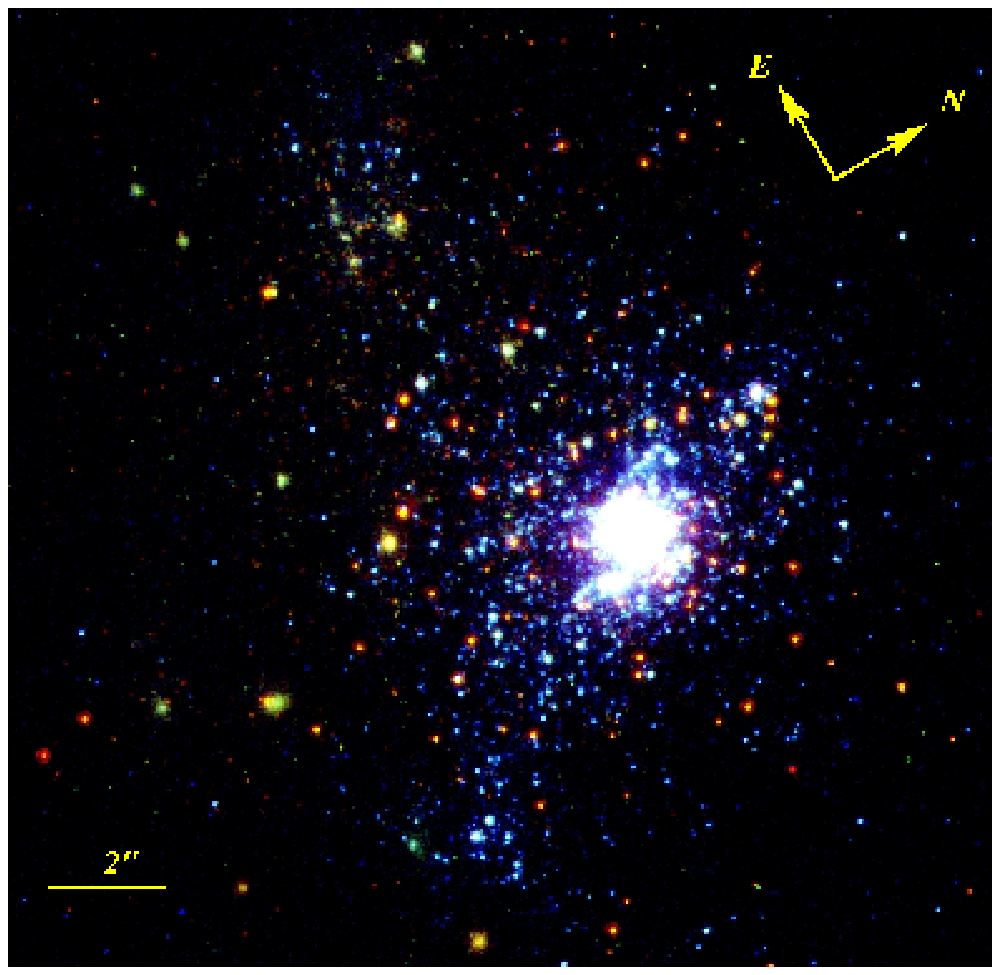}
\caption{HST image of NGC\,1705 obtained using ACS/HRC with the F330W
(blue), F555W (green), and F814W (red) filters.  Individual exposures
were downloaded from the HST data archive (PID=9989) and processed
using standard pipeline techniques.  The processed frames were stacked
using our own software.  The super star cluster, made of primarily
young blue stars, appears as a compact core near the center.}
\end{center}
\end{figure}

\begin{figure}
\begin{center}
\includegraphics[scale=0.7,angle=270]{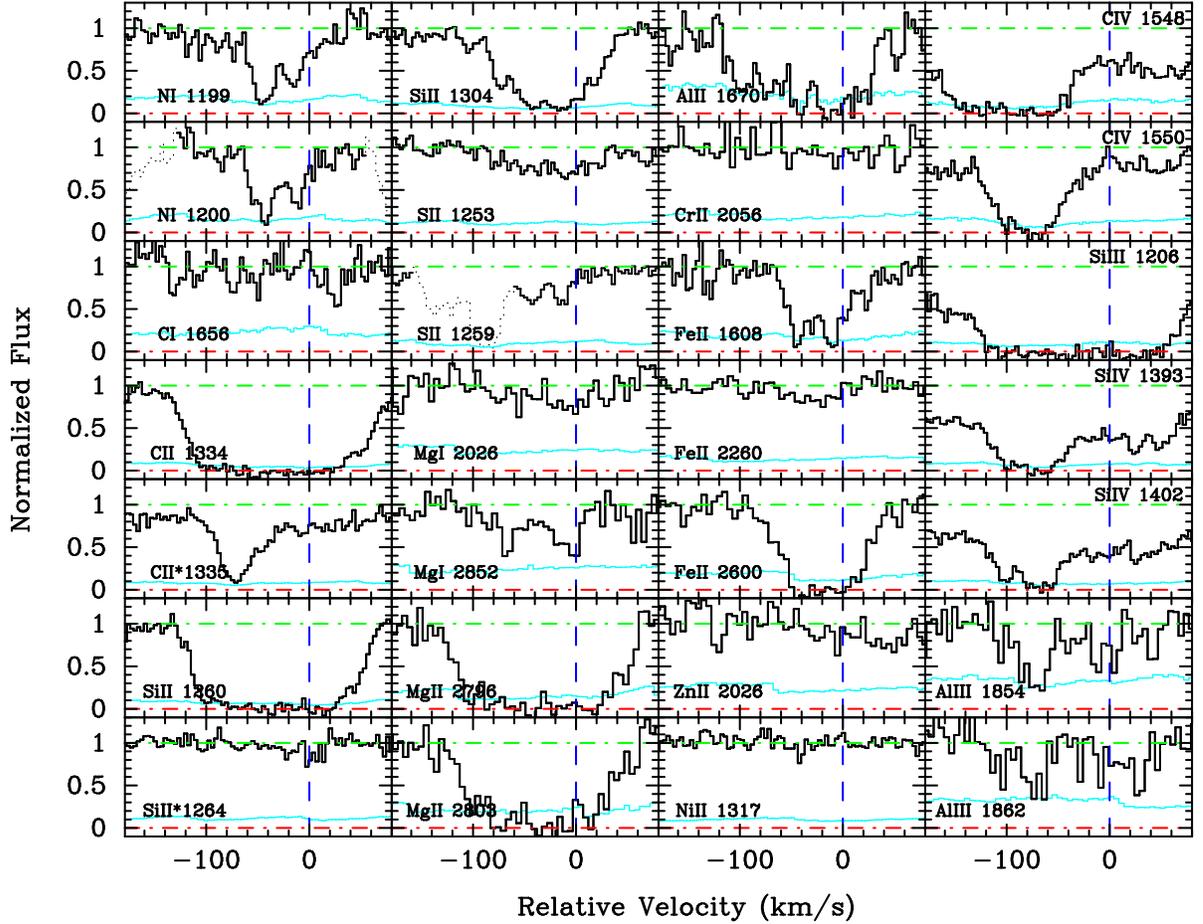}
\caption{Complex absorption profiles of heavy ions identified at the
velocity of NGC\,1705 in the HST/STIS echelle spectrum.  The three left
columns show low ionization species, and the right column shows
high-ionization species.  The zero relative velocity ($v=0$)
corresponds to $z=0.00209$.  Contaminating features are dotted out.
Stellar photospheric features are present in transitions such as C\,IV
and Si\,IV, as a broad, shallow component extending from
$v=-500$ \kms\ through $v=+300$ \kms.  While strong
low-ionization transitions such as C\,II 1334 and Si\,II 1260 are
saturated, weaker transitions such as N\,I, S\,II 1259, and Fe\,II
1608 clearly exhibit two narrow components in the neutral ISM at
$v=-15$ and $-45$ \kms.  High ionization transitions such as
Al\,III, CIV, and Si\,IV exhibit a dominant component that is
systematically shifted to $v=-70$ \kms, coincident with the
kinematic signatures of the O\,VI absorption doublet discussed in
Heckman \etal\ (2001).  We also note a weak contribution from highly
ionized gas at $v > 0$ \kms, presumably from photospheric absorption. }

\end{center}
\end{figure}

\begin{figure}
\begin{center}
\includegraphics[scale=0.6, angle=90]{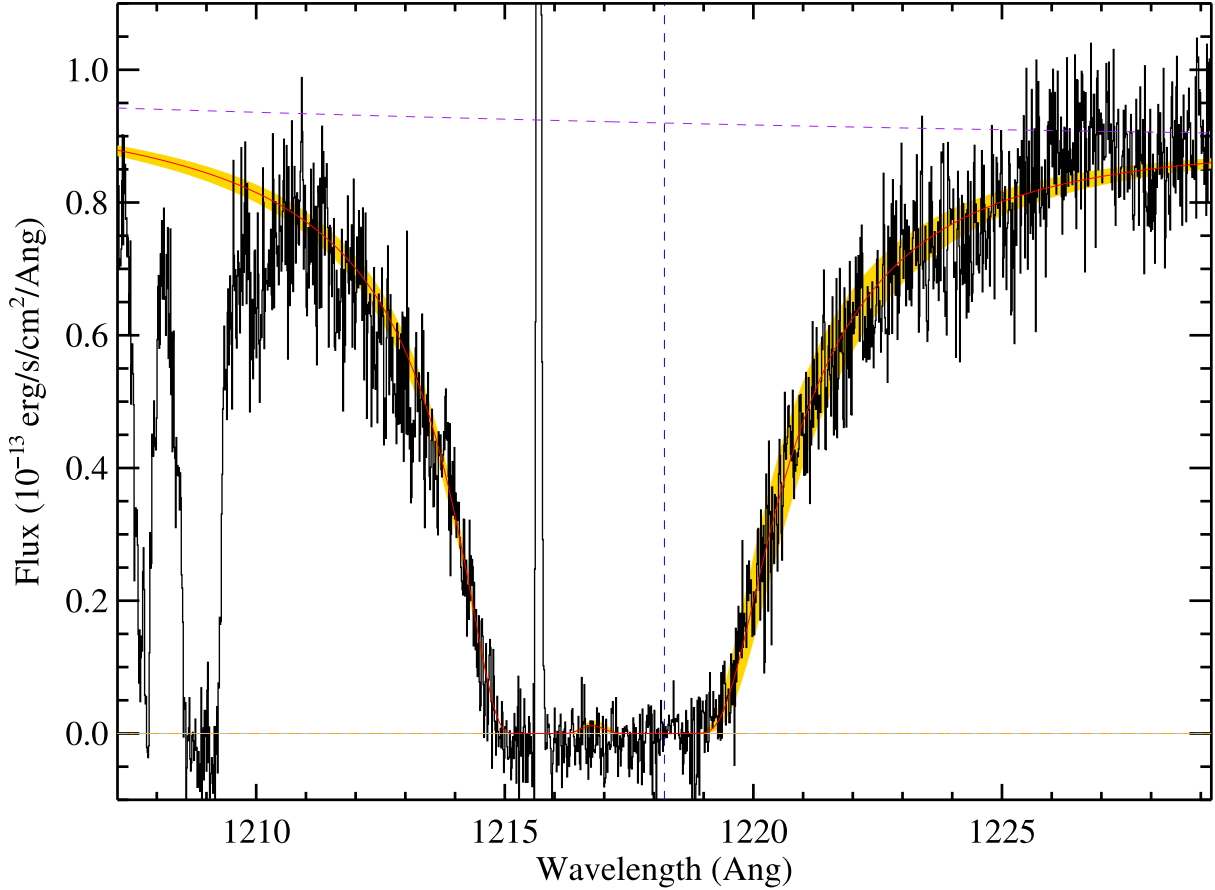}
\caption{Absorption profile of the hydrogen \lya\ transition toward
NGC\,1705-1.  The black histogram represents the echelle spectrum and
the thin curve at the bottom represents the corresponding 1-$\sigma$
array.  A Voigt profile analysis that includes two H\,I components at
$z=0.00209$ (the position of the vertical dashed line) and $z=0$ (the
emission feature) yield $\log\,N(\hI)=20.05\pm 0.15$ for NGC\,1705 and
$\log\,N(\hI)=19.7$ for the Milky Way.  The best-fit model is
represented by the red curve with the yellow band marking the
1-$\sigma$ uncertainties in the NGC~1705 \lya\ profile.  The error is
dominated by systematic uncertainty due to continuum placement which
is described by the nearly horizontal dashed line.  Although the
absorption feature is blended with the H\,I \lya\ absorption from the
Milky Way, the total neutral hydrogen absorption column density
$N(\hI)$ is well constrained by the red damping wind.  Our $N(\hI)$
measurements agree very well with previous measurements by Sahu (1998)
and Heckman \etal\ (2001) based on HST GHRS and FUSE observations,
respectively.  We note that the absorption features at $\lambda <
1210$ \AA\ are due to Si\,III 1206 transitions at $z\le 0.00209$.}
\end{center}
\end{figure}

\begin{deluxetable}{p{3in}cl}
\tabletypesize{\normalsize}
\tablewidth{0pt} 
\tablecaption{Summary of Previously Derived Stellar and ISM Properties}
\tablehead{\colhead{Property} & \colhead{Measurement} & \colhead{References}}
\startdata  
\multicolumn{3}{c}{Galaxy NGC\,1705} \nl
\hline
Distance (Mpc) \dotfill & $5.1\pm 0.6$ & Tosi \etal\ (2001) \nl
$M_B$ \dotfill & $-15.6\pm 0.2$ & Marlowe \etal\ (1999) \nl
Star formation rate ($M_\odot$ yr$^{-1}$) \dotfill & 0.06 & Lee \& Skillman (2004) \nl
Formation time scale ($\tau=L_B/{\rm SFR}$ Gyr) \dotfill & $4.3$ & Lee \& Skillman (2004) \nl
\hline
\multicolumn{3}{c}{The Star Cluster, NGC\,1705-1} \nl
\hline
Stellar Mass ($10^5\,M_\odot$) \dotfill &  $(4.8\pm 1.2)$ &  Smith \& Gallagher (2001) \nl
Effective Radius (pc) \dotfill & $1.6\pm 0.4$ & Smith \& Gallagher (2001) \nl
Age (Myr) \dotfill & $12^{+3}_{-1}$ & V\'azquez \etal\ (2004) \nl
H\,II region $12+\log ({\rm O}/{\rm H})$ \dotfill & $8.21\pm 0.05$ & Lee \& Skillman (2004) \nl
H\,II region $12+\log ({\rm N}/{\rm H})$ \dotfill & $6.46\pm 0.08$ & Lee \& Skillman (2004) \nl
$T_e$ ($10^4$ K) \dotfill & 1.1--2 & Lee \& Skillman (2004) \nl
$E(B-V)$ \dotfill & $\apl\, 0.02$ mag & Calzetti \etal\ (1994) \nl
\hline
\multicolumn{3}{c}{The Neutral ISM} \nl
\hline
$v$ (\kms) \dotfill & $-43$ & V\'azquez \etal\ (2004) \nl
$N({\rm H}_2)$ (cm$^{-2}$) \dotfill & $<\,3.9\times 10^{14}$ & Hoopes \etal\ (2004) \nl
$N(\hI)$ (cm$^{-2}$) (21\,cm) \dotfill & $(1-2)\times 10^{21}$ & Meurer \etal\ (1998) \nl
$\log\,N(\hI)$ (\lyb\ absorption) \dotfill & $20.2\pm 0.2$ & Heckman \etal\ (2001) \nl
$\log\,N({\rm O\,I})$ \dotfill & $15.63\pm 0.08$ & Heckman \etal\ (2001)\tablenotemark{a} \nl
$\log\,N(\nI)$ \dotfill & $13.97\pm 0.08$ & Heckman \etal\ (2001) \nl
$\log\,N({\rm Si\,II})$ \dotfill & $14.61\pm 0.14$ & Heckman \etal\ (2001) \nl
$\log\,N({\rm Fe\,II})$ \dotfill & $14.54\pm 0.10$ & Heckman \etal\ (2001) \nl
\hline 
\multicolumn{3}{c}{The Warm Photo-ionized ISM} \nl 
\hline 
$v$(\kms) \dotfill & $-77$ & V\'azquez \etal\ (2004) \nl 
$\log\,N({\rm N\,II})$ \dotfill & $14.70\pm 0.15$ & Heckman \etal\ (2001) \nl
$\log\,N({\rm O\,VI})$ \dotfill & $14.26\pm 0.08$ & Heckman \etal\ (2001) \nl 
\hline 
\enddata 
\tablenotetext{a}{Given the observed column density, we expect the
line to be saturated.  The reported value should therefore be treated
as a lower limit.}

\end{deluxetable}

\begin{deluxetable}{lccccccccc}
\tablewidth{0pc} \tablecaption{IONIC COLUMN DENSITIES FROM STIS OBSERVATIONS
\label{tab:colm}} \tabletypesize{\footnotesize}
\tablehead{\colhead{Ion} & \colhead{$J^a$} & \colhead{$E_{J}$} &
\colhead{$\lambda$} & \colhead{$\log f$} & \colhead{$v_{int}^b$} &
\colhead{$W_\lambda^c$} & \colhead{$\log N$} & \colhead{$\log
N_{adopt}$} \\ 
& & (cm$^{-1}$) & (\AA) & & (\kms) & (m\AA) }
\startdata 
\ion{C}{1}\\ &0 & 0.00 & 1277.2450 &$ -0.8816$&$[-100,
40]$&$< 18.4$&$<13.17$&$< 13.17$ \\ 
&0 & 0.00 & 1560.3092 &$-0.8808$&$[-100, 40]$&$< 39.6$&$<13.51$&\\ 
&0 & 0.00 & 1656.9283 &$-0.8273$&$[-100, 40]$&$< 58.7$&$<13.44$&\\ 
\ion{C}{2}\\ 
&1/2 & 0.00 & 1334.5323 &$ -0.8935$&$[-100, 40]$&$ 625.4 \pm 4.4$&$>15.13$&$> 15.13$\\ 
&3/2 & 63.40 & 1335.7077 &$ -0.9397$&$[-100, 40]$&$ 276.8 \pm 7.6$&$14.39 \pm 0.02$&$ 14.39 \pm 0.02$\\ 
\ion{C}{4}\\ 
& & 0.00 & 1548.1950 &$ -0.7194$&$[-250, 100]$&$1139.5 \pm 22.1$&$>14.85$&$> 14.88$\\ 
& & 0.00 & 1550.7700 &$ -1.0213$&$[-250, 100]$&$ 698.1 \pm 24.0$&$>14.88$&\\ 
\ion{O}{1}\\ &2 & 0.00 & 1302.1685 &$ -1.3110$&$[-60, 40]$&$ 366.0 \pm 4.8$&$>15.24$&$> 15.24$\\ 
&1 & 158.26 & 1304.8576 &$ -1.3118$&$[-100, 40]$&$ 55.2 \pm 8.8$&$14.05 \pm 0.06$&$14.05 \pm 0.06$\\ 
\ion{Mg}{1}\\ 
& & 0.00 & 2852.9642 &$0.2577$&$[-100, 40]$&$ 349.3 \pm 64.3$&$12.54 \pm 0.09$&$ 12.54 \pm 0.09$\\ 
\ion{Mg}{2}\\ & & 0.00 & 2796.3520 &$ -0.2130$&$[-100,40]$&$1269.5 \pm 34.3$&$>13.97$&$> 14.14$\\ 
& & 0.00 & 2803.5310 &$-0.5151$&$[-100, 40]$&$1146.5 \pm 49.3$&$>14.14$&\\ 
\ion{Al}{2}\\ 
& & 0.00 & 1670.7874 &$ 0.2742$&$[-100, 40]$&$ 585.8 \pm 23.0$&$>13.49$&$>13.49$\\ 
\ion{Al}{3}\\ 
& & 0.00 & 1854.7164 &$ -0.2684$&$[-100, 40]$&$274.4 \pm 52.1$&$13.39 \pm 0.10$&$ 13.43 \pm 0.08$\\ 
& & 0.00 &1862.7895 &$ -0.5719$&$[-100, 40]$&$ 200.1 \pm 54.4$&$13.53 \pm 0.13$&\\ 
\ion{Si}{2}\\ 
&1/2 & 0.00 & 1260.4221 &$ 0.0030$&$[-100,40]$&$ 583.6 \pm 5.6$&$>14.21$&$> 14.83$\\ 
&1/2 & 0.00 & 1304.3702 &$-1.0269$&$[-100, 40]$&$ 401.8 \pm 6.6$&$>14.81$&\\ 
&1/2 & 0.00 &1526.7066 &$ -0.8962$&$[-100, 40]$&$ 574.5 \pm 11.0$&$>14.83$&\\ 
&3/2 & 287.24 & 1264.7377 &$ -0.0441$&$[-100, 40]$&$< 19.9$&$<12.39$&$< 12.39$\\
\ion{S}{2}\tablenotemark{d} 
\\ & & 0.00 & 1259.519 & $-1.7894$ & ... & ... & $14.67\pm 0.10$ & $14.65\pm 0.09$ \\
\ion{Cr}{2}\\ 
& & 0.00 & 2056.2539 &$ -0.9788$&$[-100, 40]$&$<65.0$&$<13.42$&$< 13.42$\\ 
\ion{Fe}{2}\\ 
&9/2 & 0.00 & 1608.4511 &$-1.2366$&$[-100, 40]$&$ 307.6 \pm 19.0$&$14.64 \pm 0.06$&$ 14.64 \pm 0.06$\\ 
&9/2 & 0.00 & 1611.2005 &$ -2.8665$&$[-100, 40]$&$< 50.8$&$<15.60$&\\ 
&9/2 & 0.00 & 2260.7805 &$ -2.6126$&$[-100, 40]$&$< 55.5$&$<14.91$&\\ 
&9/2 & 0.00 & 2344.2140 &$ -0.9431$&$[-100, 40]$&$575.9 \pm 24.8$&$>14.33$&\\ 
&9/2 & 0.00 & 2374.4612 &$-1.5045$&$[-100, 40]$&$ 379.8 \pm 39.5$&$>14.65$&\\ 
&9/2 & 0.00 &2382.7650 &$ -0.4949$&$[-100, 40]$&$ 737.3 \pm 37.7$&$>14.08$&\\
\ion{Ni}{2}\\ 
& & 0.00 & 1370.1310 &$ -1.2306$&$[-100, 40]$&$<21.6$&$<13.53$&$< 13.53$\\ 
\ion{Zn}{2}\\ 
& & 0.00 & 2026.1360 &$-0.3107$&$[-100, 40]$&$< 85.7$&$<12.90$&$< 12.90$\\ 
\enddata
\tablenotetext{a}{Total angular momentum of the electron spin and
orbital angular moment.  $E_{J}$ is the energy above the ground
state.}  
\tablenotetext{b}{Velocity interval over which the equivalent
width and column density are measured.}  
\tablenotetext{c}{Rest equivalent width.}  
\tablenotetext{d}{The S\,II 1259 transition is
contaminated by the Si\,II 1260 transition from a foreground
high-velocity cloud along the sightline.  We determine $N({\rm
S\,II})$ using the VPFIT software package, including multiple
components.  Here we present the sum of the two narrow components at
$\Delta\,v=-45$ and $-15$ \kms\ (see Figure 1).}
\end{deluxetable}

\begin{deluxetable}{p{1.5in}cc}
\tabletypesize{\normalsize}
\tablewidth{0pt} 
\tablecaption{Comparison of ISM Properties}
\tablehead{\colhead{Property} & \colhead{NGC\,1705-1} & \colhead{GRB Hosts}}
\startdata  
$\log\,N(\hI)$ \dotfill & $[20.1,21.2]$ & $[19,8,22.5]$ \nl 
$[{\rm Z}/{\rm H}]$\tablenotemark{a} \dotfill &  $-0.6\pm 0.1$ &  $[-2.2,-0.2]$ \nl
$[{\rm \alpha}/{\rm Fe}]$\tablenotemark{a} \dotfill &  $+0.5\pm 0.2$ &  $[+0.4,+0.9]$ \nl
$[{\rm N}/{\rm \alpha}]$ \tablenotemark{a}\dotfill &  $-1.5\pm 0.2$ &  $[-2.0,-0.5]$ \nl
$f_{\rm H_2}$\tablenotemark{b} \dotfill & $< 5.2\times 10^{-7}$ & $< 3\times 10^{-7}$ \nl
$A_V$\tablenotemark{a} \dotfill & $\sim 0.0$ & $[0.0,0.2]$ \nl
$d_{\rm neutral}$ (pc) \dotfill & $\apl 500$ & 100--1000 \nl
\enddata
\tablenotetext{a}{Measurements for the ISM of GRB hosts are from Prochaska \etal\ (2007). }
\tablenotetext{b}{Measurements for GRB hosts are from Tumlinson \etal\ (2007). }
\end{deluxetable}

\end{document}